\RequirePackage{fix-cm}

\documentclass[a4paper,10pt,pdftex]{article}
\usepackage{fullpage}

\usepackage{authblk}
\usepackage{graphicx}
\usepackage{amsmath}
\usepackage{amsfonts}
\usepackage{amssymb}
\usepackage{algorithm}
\usepackage[noend]{algorithmic}
\usepackage[numbers,sort&compress,square]{natbib} 
\usepackage[utf8]{inputenc}    
\usepackage[T1]{fontenc}       
\usepackage{tikz}
\usepackage{multirow}
\usepackage{booktabs}
\usetikzlibrary{positioning,calc}
\usetikzlibrary{fit}
\usepackage{booktabs}
\usepackage{soul}
\usepackage[normalem]{ulem}

\newlength\myindent
\newcommand\bindent{%
  \begingroup
  \setlength{\itemindent}{\myindent}
  \addtolength{\algorithmicindent}{\myindent}
}
\newcommand\eindent{\endgroup}

\begin{document}
%
\title{A novel algorithm for online inexact string matching and its FPGA implementation}

\author[1]{Alessandro Cinti}
\author[2]{Filippo Maria Bianchi}
\author[1]{Alessio Martino}
\author[1]{Antonello Rizzi}
\affil[1]{Department of Information Engineering, Electronics and Telecommunications, University of Rome "La Sapienza", Via Eudossiana 18, 00148 Rome, Italy}
\affil[2]{Machine Learning group, Department of Physics and Technology, UiT the Arctic University of Norway, Hansine Hansens veg 18, 9019 Troms{\o}, Norway}
\setcounter{Maxaffil}{0}
\renewcommand\Affilfont{\itshape\small}

\date{}
\maketitle

\begin{abstract}
Accelerating inexact string matching procedures is of utmost importance when dealing with practical applications where huge amount of data must be processed in real time, as usual in bioinformatics or cybersecurity. 
Inexact matching procedures can yield multiple shadow hits, which must be filtered, according to some criterion, to obtain a concise and meaningful list of occurrences. 
The filtering procedures are often computationally demanding and are performed offline in a post-processing phase.
This paper introduces a novel algorithm for Online Approximate String Matching (OASM) able to filter shadow hits on the fly, according to general purpose priority rules that greedily assign priorities to overlapping hits. 
An FPGA hardware implementation of OASM is proposed and compared with a serial software version. 
Even when implemented on entry level FPGAs, the proposed procedure can reach a high degree of parallelism and superior performance in time compared to the software implementation, while keeping low the usage of logic elements. 
This makes the developed architecture very competitive in terms of both performance and cost of the overall computing system.
\end{abstract}


\section{Introduction}

Approximate String Matching (ASM) studies the problem of matching two generic strings defined on the same alphabet and, unlike in Exact String Matching, a level of inexactness is allowed: two strings match if their dissimilarity is below a given threshold \cite{navarro2001guided}. 
ASM is a particular case of subgraph matching~\cite{livi2013graph,Tran2016} that has been applied in different fields of science and technology, including computational biology, signal processing and text retrieval. 
In computational biology, scientists try to detect frequent patterns into DNA sequences of nucleotides (motifs) using ASM to account for mutations and evolutionary alterations of the genome sequences \cite{buhler2002finding,eskin2002finding,pavesi2004weeder,sinha2003ymf}.
In signal processing, ASM can be used to identify patterns which have been corrupted by noise during storage, transmission, processing or conversion stages. 
The main application fields are message exchanges, wireless communications, audio, image and video processing \cite{typke2005survey, bertini2006video}. 
ASM is extensively used for information retrieval in large text collections where, due to the large amount of data and the high variety of content, classical string matching procedures are usually not suitable \cite{ziviani2000compression, Boukharouba2011, Sahi2017, gravano2003text, maiorino2016noise}.

ASM also plays a fundamental role in Granular Computing, which is a human-inspired computing and information processing paradigm that explores multiple levels of granularity in data~\cite{yao2008rise}. 
The concept of Granular Computing arose from many branches of natural and social sciences~\cite{Howard2014, bargiela2016granular, yao2016triarchic} and it is at the basis of recently developed frameworks in computational intelligence~\cite{Singh2018, lin2013data, bianchi2014granular}.
In this context, an ASM technique that identifies frequent and meaningful motifs in sequences database allows to design advanced machine learning systems such as Symbolic Histograms approaches~\cite{rizzi2012new, Bianchi2016, Martino2018}, where each pattern (a sequence of objects/events) can be represented by a histogram of motif instances.

The dissimilarity between two strings can be evaluated as the cost of the edit operations required for transforming a string into the other. The most common dissimilarities are the Levenshtein, Hamming, Episode, and Longest Common Subsequence distance; they differ on the edit cost definition and on the type of the allowed edit operations, and their computational costs range from linear to NP-complete~\cite{andoni2010polylogarithmic}. 
The matching procedure can be implemented following offline or online approaches, which differ on how sequences are searched and indexed~\cite{boytsov2011indexing}. 
Several algorithms for online pattern matching have been designed and they can be grouped in four main classes, namely the Dynamic Programming (DP), Automata, Bit-parallel, and Filtering approach \cite{navarro2001guided}.
The main advantage of online ASM is to provide detection in real time, which is essential in situations where a prompt response is required.

Cybersecurity is a prominent example of application of online ASM.
In particular, Network Intrusion Detection Systems (NIDSs) are devices or software applications used to identify individuals who are using a computer system without authorisation and whoever has legitimate access to the system, but is exceeding his privileges~\cite{di2008intrusion}.
Traditional NIDSs relied on exact pattern matching to detect an attack.
However, by changing the data used in the attack even slightly, it is possible to evade detection. 
Therefore, a more flexible detection system is required to scan efficiently \textit{in real-time} the whole inbound and outbound traffic to match patterns from a library of known attacks, without compromising the overall network speed.

In molecular biology, online ASM is exploited in quantitative real-time Polymerase Chain Reaction (PCR) \cite{heid1996real}. 
Real-time PCR aims at amplifying a small target DNA sample during PCR, performing a quantitation step after amplification. 
This technique is useful when only a small amount of DNA is available, which is insufficient for performing an accurate analysis. 
In medicine, real-time PCR is applied to the discovery of tumour cells, to the diagnosis of infectious diseases, and to a plethora of other predictive medicine and diagnostic tasks \cite{espy2006real}; it is widely used for studying genomes in several bacteria and protists which cannot be cultured; in legal medicine and forensics it has been demonstrated to be crucial for analysing fingerprints and stains found at crime scenes \cite{MADEL2016166,NIEDERSTATTER20061}.

To increase the time performance of string matching procedures, especially when dealing with big-data, their concurrent processes can be effectively parallelized at CPU level by executing microinstructions simultaneously, or at circuit level by re-implementing arithmetic and logic operations \cite{rasool2012parallelization, zhong2007fast}.
Hardware implementation provides an alternative solution to improve the speed of ASM algorithms and several architectures have been proposed. 
The fastest implementations are custom architectures deploying a large number of processing elements designed \textit{ad-hoc} to execute a specific algorithm~\cite{crochemore2001fast}.
Other architectures, instead, are based on general-purpose processors that can be adopted to implement different algorithms \cite{leighton2014introduction, michailidis2005programmable}. 
Configurable devices, such as the Field-Programmable Gate Array (FPGA), represent flexible hardware platforms that are used to accelerate the computation, providing both a high degree of programmability and reduced design time and costs~\cite{Antonik2017, Vasquez2013}. 

None of the existing methods for performing online ASM accounts for the presence of multiple shadow hits, which are overlapping hits that appear as a consequence to the flexibility in ASM. 
Usually, such shadow hits are filtered afterwards by means of a post-processing phase which increases the overall computational burden. 
Especially when dealing with data streams, this solution turns out to be unfeasible both in terms of memory and time complexity, hence the necessity to filter out shadow hits on the fly.

This paper introduces the Online Approximate String Matching (OASM) algorithm for retrieving a set of inexact matches of a known pattern from a stream of symbols, while filtering shadow hits on the fly.
The proposed method performs detection in real time and is suitable for real-time applications and big-data streams. 
OASM is mainly based on the evaluation of the Levenshtein distance between a target pattern and a set of sequences, obtained by shifting windows of pre-established lengths over the stream. 
OASM follows a DP approach that provides a remarkable speedup thanks to the massive parallelization capability deriving from the inherent simplicity and regularity of its structure. 
Both a software (SW-OASM) and a hardware (HW-OASM) implementation of the algorithm are proposed in this paper. 
HW-OASM is based on the systolic arrays principle for the computation of the Levenshtein distance that extends the work in~\cite{mikami2008efficient}.
Compared to its software counterpart, experimental results show the effectiveness of the HW-OASM especially when integrated in a Multiple Online Approximate String Matching (MOASM) system.
The latter, implements multiple instances of HW-OASM to perform simultaneously parallel searches of distinct patterns on a common input stream of symbols (not necessarily finite). The performance of the proposed algorithm is first evaluated in a controlled environment using synthetic data.
Successively, the algorithm is applied to a case study in bioinformatics on real genome data. The analysis consists in mining part of the human genome for RNA-binding protein sites.

The remainder of the paper is organized as follows. Sec.~\ref{sec:relatedWorks} reviews related ASM approaches, including hardware-based ones. Sec.~\ref{sec:definitions} introduces basic concepts and details of the dissimilarity measure considered. 
Sec.~\ref{sec:algo} describes the OASM algorithm and in Sec.~\ref{sec:casestudy} a real-world human genome case study is presented to highlight the features of SW-OASM, compared to other ASM approaches. 
Next, Sec.~\ref{sec:hardware} provides details of the hardware implementation and in Sec.~\ref{sec:experiments} the performances of SW-OASM and HW-OASM are compared on synthetic data. 
Finally, in Sec. \ref{sec:concl} conclusions are reported.

\section{Related Works}\label{sec:relatedWorks}
In literature, there are two major approaches based on DP for performing ASM on dedicated hardware, namely FPGAs and (GP-)GPUs. 
In this paper we focus on FPGAs and refer the interested reader to works such as \cite{7797444,XU2013523,7424709,10.1371/journal.pone.0186251} for ASM implementations on (GP-)GPUs. 
The first approach relies on local/global sequence alignment \cite{van2004families} (e.g. the Smith-Waterman's algorithm \cite{SMITH1981195}, the Needleman-Wunsch's algorithm \cite{needleman1970general} or Myers' fast bit-vector algorithm \cite{myers1999fast}), while the second approach exploits the Levenshtein distance (e.g. the Wagner-Fischer's algorithm \cite{wagner1974string}).

Implementation of the Smith-Waterman's algorithm for optimal local alignment on FPGAs can be found in \cite{yu2005smith}, where the authors implement the plain Smith-Waterman's algorithm. In \cite{dydel2004large} the Smith-Waterman's algorithm has been implemented on FPGAs without relying on systolic arrays in order to exploit all processing units, whereas in \cite{sirasao2015fpga} the authors used systolic arrays driven by OpenCL. In \cite{herbordt2006single} the authors provide FPGA implementation for DP and BLAST routines \cite{ALTSCHUL1990403}. It is noteworthy that since the Smith-Waterman's algorithm was primarily developed for nucleotide or protein sequence alignment, all works cited so far regard bioinformatics-related applications. Conversely, in \cite{west2003fpga}, the authors implement the Smith-Waterman's algorithm on FPGA for generic text search. Finally, in \cite{10.1007/978-3-319-32703-7_104}, the authors use an FPGA in order to speedup Myers' algorithm.

General purpose applications are more frequently relying on the second approach, based on the Wagner-Fischer's algorithm. In \cite{mikami2008efficient} the authors adopt the plain Wagner-Fischer's algorithm, along with an improved version that better exploits each cell. In \cite{bluthgen2000programmable} the Wagner-Fischer's algorithm has been applied to multimedia information retrieval, also considering text search paradigms such as wildcards and idioms. In \cite{utan2010fpga} the authors provide an FPGA implementation of the Wagner-Fischer's algorithm particularly suited for dealing with regular expressions.

FPGA implementations of systolic architectures have been proposed for ASM \cite{mikami2008efficient, west2003fpga} and, in bioinformatics, for DNA sequence alignment \cite{yu2005smith, dydel2004large, herbordt2006single}. 
In the field of Music Information Retrieval (MIR), circuits implementing ASM with dynamic programming cannot retrieve fragments with different sizes  \cite{bluthgen2000programmable} and other architectures with higher flexibility only support symbol substitution \cite{park1999parallel}. 
Better performance was achieved by Application-Specific Integrated Circuit and FPGA with a comparable computational time on different MIR approaches \cite{ou2008fpga, smith2008application, brown1996fpga, bondalapati2002reconfigurable}.
An efficient FPGA implementation of ASM has been proposed for text mining, where a restricted class of regular expressions is used to define the patterns to search~\cite{utan2010fpga}.
A work closely related to this paper is a successful implementation of online ASM on FPGA for NIDS applications~\cite{kawanaka2008systolic}.

Regardless of the particular algorithm used to implement ASM (online and/or using dedicated hardware), none of the works previously discussed filters shadow hits on the fly. In fact, those are discarded {\em a-posteriori}, increasing the overall computational burden. 
The proposed OASM jointly provides (i) the ability of performing simultaneously detection and shadow hits filtering; (ii) the ability of working with data streams, avoiding to store the entire input sequence and/or all multiple shadow hits. 
Contrarily to previous works, in OASM not only the plain Levenshtein distance is parallelized, but also the filtering procedure.
This makes the proposed approach a perfect candidate for being implemented in FPGAs.


\section{An Overview on the Levenshtein Distance}
\label{sec:definitions}
Let the pattern $\mathbf{p} = \langle p^{(0)}, p^{(1)}, \dots, p^{(l_p - 1)} \rangle$ and the string $\mathbf{t} = \langle t^{(0)}, t^{(1)}, \dots, t^{(l_t - 1)} \rangle$ be defined, respectively, as the concatenation of $l_p$ and $l_t$ symbols of a finite alphabet $\Sigma$ and a generic substring $\mathbf{s} = \langle s^{(0)}, s^{(1)}, \dots, s^{(l-1)} \rangle = \langle t^{(i)}, t^{(i+1)}, \dots, t^{(i+l-1)} \rangle = \mathbf{t}[i, l]$ be a subset of $l$ contiguous symbols in $\mathbf{t}$ starting from position $i$.
The dissimilarity between strings $\mathbf{p}$ and $\mathbf{s}$ can be measured by means of the Levenshtein distance $\mathrm{lev}(\mathbf{p}, \mathbf{s}) \in \mathbb{N}_0$, which is the minimum number of single-character edits (insertion, deletion, substitution) necessary to transform $\mathbf{p}$ into $\mathbf{s}$. 
Let $\mathbf{C}$ be a matrix of size $(l_p+1) \times (l+1)$, whose elements are 
\begin{align}
\label{eq:lev_mat}
c_{n,m} = \begin{cases} n, & \mbox{if } m = 0 \\ m, & \mbox{if } n = 0 \\ \min( c_{n-1,m}+1, c_{n,m-1}+1, \\ c_{n-1,m-1} + \delta_{n,m}), & \mbox{if } n,m > 0 \end{cases}
\end{align}
where $\delta_{n,m} = \left( p^{(n)} \neq s^{(m)} \right)$, $n = 0, \dots, l_p$, $m = 0, \dots, l$.
The Levenshtein distance between $\mathbf{p}$ and $\mathbf{s}$ corresponds to the element $c_{l_p,l}$ of the Levenshtein matrix $\mathbf{C}$.
Eq.~\ref{eq:lev_mat} shows that, for $n,m > 0$, each element $c_{n,m}$ of the matrix $\mathbf{C}$ can be computed just by knowing its upper ($c_{n-1,m}$), left ($c_{n,m-1}$), and upper-left ($c_{n-1,m-1}$) neighbours. 
Applying the DP method to the Levenshtein distance computation consists in building column-wise (or row-wise) the matrix $\mathbf{C}$, element by element, by solving iteratively the same simple problem, whose result will be used as input of one of the successive iterations. Although the algorithm is $\mathcal{O}\left(l_p \cdot l\right)$ in time, the space complexity is only $\mathcal{O}\left(\min(l_p,l)\right)$ because only the previous column (or row) has to be stored to compute the new one. 

Applying ASM between $\mathbf{p}$ and $\mathbf{t}$ consists in finding the set of all the substrings $\mathbf{s} \subset \mathbf{t}$ with length $l_s \in [l_p-K, l_p+K]$ that satisfy the condition
\begin{equation}
\label{eq:inexact}
    \mathrm{lev}(\mathbf{p}, \mathbf{s}) = k \leq K, 
\end{equation}
where the threshold $K$ represents the maximum acceptable level of inexactness~\cite{levenstein1965binary}.
\begin{figure}[!ht]
\centering
\includegraphics[width=0.65\columnwidth, keepaspectratio,trim={0cm 0cm 0cm 0cm},clip]{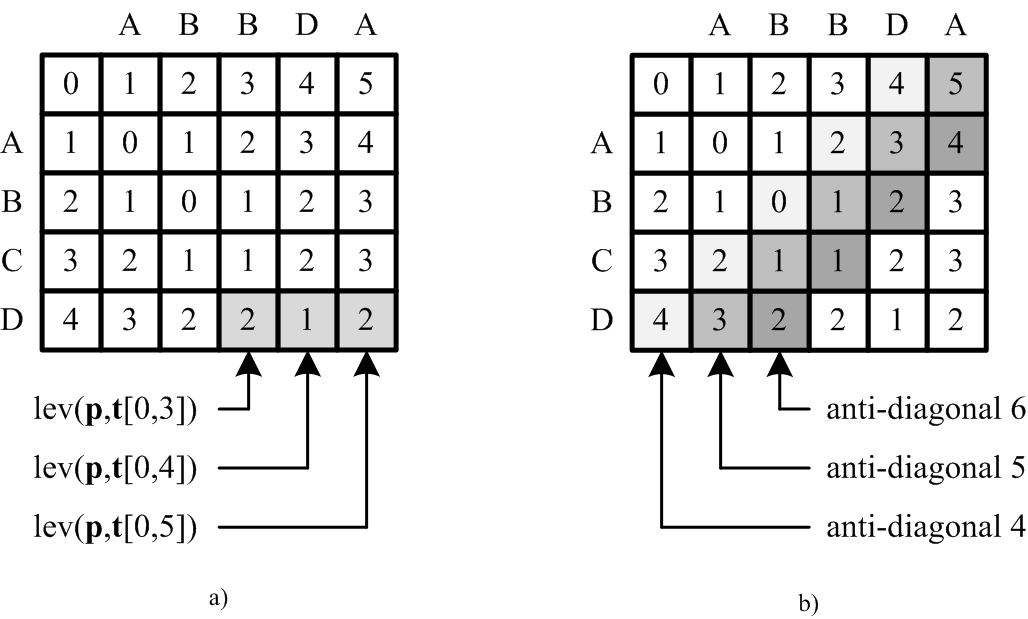}
\caption{Example of the matrix $\mathbf{C}$ computation with canonical (a) and wave-front (b) approach.}
	\label{fig:fig1}
\end{figure}
Fig.~\ref{fig:fig1}(a) depicts the matrix $\mathbf{C}$ resulting from the computation of the Levenshtein distance between the pattern $\mathbf{p} = \langle ABCD \rangle$ and the substring $\mathbf{s} = \langle ABBDA \rangle$ from the string $\mathbf{t} = \langle ABBDABCDACDB \rangle$ defined over the alphabet $\Sigma = \left\{ A, B, C, D \right\}$ for $K = 1$ and $i = 0$. 
The gray-shaded cells in Fig.~\ref{fig:fig1}(a) represent the values of the Levenshtein distances between $\mathbf{p}$ and the three substrings $\mathbf{t}[0,3] = \langle ABB \rangle$, $\mathbf{t}[0,4] = \langle ABBD \rangle$, and $\mathbf{t}[0,5] = \langle ABBDA \rangle$. 
There is no need to compute three different distances since by computing only the distance between $\mathbf{p}$ and $\mathbf{t}[0,5]$ (element $c_{4,5}$ of $\mathbf{C}$), the other two distances are available as intermediate computations (elements $c_{4,3}$ and $c_{4,4}$ of $\mathbf{C}$, respectively). 
By increasing $i$, step by step, all the substrings of $\mathbf{t}$ are generated. 

The DP method can be applied to compute subsequently the Levenshtein distances between $\mathbf{p}$ and all the substrings obtained by shifting a fixed mask of length $l_p+K$ over $\mathbf{t}$~\cite{SELLERS1980359}. 
The DP approach reduces the overall number of computations compared to brute force approaches or when the same computation is repeated over and over~\cite{UKKONEN1985100}. 

The DP approach for the Levenshtein distance computation exploits the relationship between each element $c_{n,m}$ in $\mathbf{C}$ and its three neighbours $c_{n-1,m}, c_{n,m-1}, c_{n-1,m-1}$.
All the elements on an anti-diagonal $j$ can be computed at the same time in a wave-front processing fashion, by combining the information contained in the anti-diagonals $j-1$ and $j-2$, together with the information related to the symbols of the two compared strings. 
Starting from the element $c_{0,0}$ and ending with the element $c_{l_p,l}$, the matrix $\mathbf{C}$ is filled in $2 l_p+K-1$ steps, as shown in Fig.~\ref{fig:fig1}(b).
On the first step, the anti-diagonals $0$ and $1$ are combined to obtain the anti-diagonal $2$, and so on, up to the step $2 l_p+K-1$, where the anti-diagonals $2 l_p+K-1$ and $2 l_p+K$ are combined to obtain the anti-diagonal $2 l_p+K+1$.


\section{An Online Search Algorithm Based on Approximate String Matching}
\label{sec:algo}

This section presents the proposed online algorithm that combines the features of the DP method with a search criterion based on priority rules to find inexact occurrences of a known pattern $\mathbf{p}$ of length $l_p$ within a continuous stream of symbols $\mathbf{t}$.
All the found occurrences are stored in a set $\mathcal{S}$ to be used for further processing. 
Collecting all the substrings $\mathbf{s}$ of length $l \in [l_p-K, l_p+K]$ from the text $\mathbf{t}$ that verify Eq.~\ref{eq:inexact} at each position $i$, may lead to multiple hits of the same occurrence of $\mathbf{p}$~\cite{matsui2013new}.
Due to the online setting considered, it is not possible to decide whether an occurrence can be added to $\mathcal{S}$ at the time it is found, but it depends on the symbols of $\mathbf{t}$ that are not yet processed.
For instance, if Eq.~\ref{eq:inexact} is verified for two substrings $\mathbf{s'}$ and $\mathbf{s''}$ that start from two consecutive positions in $\mathbf{t}$, respectively $i$ and $i+1$, the two hits will correspond to the same occurrence of $\mathbf{p}$. 
The next subsection introduces a set of rules that determine the priorities for accepting the hits found in the text \textbf{t} as valid occurrences.

\subsection{Priority Rules}
Not all the overlapping substrings verifying Eq.~\ref{eq:inexact} should be considered as occurrences but, at the same time, the online nature of the problem complicates deciding on the fly which one should be kept or discarded.
The proposed solution consists in assigning temporary priority values to the occurrences that verify Eq.~\ref{eq:inexact}. 
In the remainder of the paper, smaller values denote higher priorities and they depend on the degree of matching between $\mathbf{p}$ and $\mathbf{s}$, and on the position of the substrings in \textbf{t}.

Three rules define a priority scheme that allows to build a greedy online algorithm with low complexity to solve the actual NP-complete problem.
\begin{description}
\item[\textbf{R1:}] Consider that the substring $\mathbf{s'} = \mathbf{t}[i',l']$ for which $\mathrm{lev}(\mathbf{p},\mathbf{s'}) = k' \leq K$ has already been found and assigned with a priority level $k'$. 
Any other overlapping substring $\mathbf{s''} = \mathbf{t}[i'',l'']$ for which $\mathrm{lev}(\mathbf{p},\mathbf{s''}) = k'' \leq K$ becomes an occurrence only if $k'' < k'$.
\item[\textbf{R2:}] If $\mathbf{s'}$ and $\mathbf{s''}$ have same priority, the first encountered substring becomes an occurrence, the other one is discarded.
\item[\textbf{R3:}] If $\mathbf{s'}$ and $\mathbf{s''}$ have same priority and start at the same position in $\mathbf{t}$, the shorter becomes an occurrence.
\end{description}

\begin{algorithm}[h!]\footnotesize
\begin{algorithmic}
\STATE \textbf{R1:}
\bindent
\IF{$\mathbf{s'} \cap \mathbf{s''} \neq \emptyset$ \AND $(k'' < k')$}
\STATE $\mathbf{s''}$ is an occurrence with priority $k''$
\ELSE
\STATE $\mathbf{s''}$ is discarded
\ENDIF
\eindent

\STATE \textbf{R2:}
\bindent
\IF{$\mathbf{s'} \cap \mathbf{s''} \neq \emptyset$ \AND $k' = k''$} 
\IF{$i' < i''$} 
\STATE $\mathbf{s'}$ is an occurrence, $\mathbf{s''}$ is discarded 
\ELSE
\STATE $\mathbf{s'}$ is discarded, $\mathbf{s''}$ is an occurrence
\ENDIF
\ENDIF
\eindent

\STATE \textbf{R3:}
\bindent
\IF {$\mathbf{s'} \cap \mathbf{s''} \neq \emptyset$ \AND $k' = k''$ \AND $i' = i''$}
\IF {$l' < l''$} 
\STATE $\mathbf{s'}$ is an occurrence, $\mathbf{s''}$ is discarded
\ELSE
\STATE $\mathbf{s'}$ is discarded, $\mathbf{s''}$ is an occurrence
\ENDIF
\ENDIF
\eindent

\end{algorithmic}
\end{algorithm}

\subsection{Validation Process}

It is possible to notice that the three rules do not resolve completely the prioritisation of the occurrences in presence of overlaps. 
For example, consider three overlapping substrings $\mathbf{s'}$, $\mathbf{s''}$, and $\mathbf{s'''}$, so that
\begin{itemize}
    \item $i' < i'' < i'''$;
    \item $k'  > k'' > k'''$;
    \item $	\mathbf{s'} \cap \mathbf{s''} \neq 0, \mathbf{s''} \cap \mathbf{s'''} \neq 0, \mathbf{s'} \cap \mathbf{s'''} = 0$
\end{itemize}
As $\mathbf{t}$ is processed symbol by symbol, priority $k' = 2$ is assigned to the occurrence relative to $\mathbf{s'}$. 
Then, the substring $\mathbf{s''}$  becomes an occurrence with priority $k'' = 1$ verifying rule R1. 
As more symbols are processed, $\mathbf{s'''}$ is found and according to rule R1 it becomes an occurrence with priority $k''' = 0$. 
Due to the condition $\mathbf{s'} \cap \mathbf{s'''} = 0$, only the occurrence relative to $\mathbf{s''}$ should be discarded.
To ensure this behaviour it is necessary to introduce a further validation procedure, which decides when it is possible to safely discard an occurrence in presence of overlapping substrings. 

The validation process proposed here uses one counter $r(k)$ per priority level:
\begin{itemize}
    \item when an occurrence with priority $k < K$ is found, its validation begins and $r(k)$ starts counting from 0 up to a value equal to the occurrence length;
    \item if overlapping occurrences with higher priority $k^* < k$ are encountered, $r(k)$ keeps counting until the end of the occurrences. 
\end{itemize}

When the validation process of the occurrence with the highest priority is completed (its relative counter $r(k)$ reaches the target value), it can be decided which of the validated occurrences can be added to $\mathcal{S}$. 
Starting from the occurrence with the highest priority and going down to the validated occurrence with the lowest priority, the substring relative to the highest priority is added to $\mathcal{S}$.
Moving downwards, the validated occurrence with lower priority $k$ is added to $\mathcal{S}$ if the following condition is verified
\begin{equation}
    \label{eq:3}
    r(k) - l(k^*) > l(k)
\end{equation}
where $k^* < k$ represents the index (priority) of the occurrence relative to the last substring added to $\mathcal{S}$.

The pseudo-code in Alg.~\ref{alg:osam} summarises the whole OASM procedure.
The storage variable \textbf{mem} is a matrix of size $(K+1) \times 3$, whose content is reset by calling ``$\mathrm{reset}(\mathbf{mem})$''.
\textbf{mem} stores in his columns three quantities for each one of the $K+1$ possible priority levels of the occurrence: its position $i$ (column 1), its length $l$ (column 2), and its associated validation counter $r$ (column 3). 
The variable \textit{idx} contains the priority of last found occurrence and the variable \textit{ins} is a flag that enables the counting check in the validation process. 
The counters $r(k)$ of the occurrences in \textbf{mem} are incremented by one each time a new symbol of \textbf{t} is processed. 
Finally, the variable \textit{acc} is used to implement Eq.~\ref{eq:3} and accumulates all the lengths $l$ of the validated occurrence that has just been added to $\mathcal{S}$. 
Every time substrings are added to $\mathcal{S}$, the variables \textit{idx}, \textit{ins}, \textit{acc}, and \textbf{mem} are reset. 
The symbol ``:'' selects all the indexes across one dimension.

\begin{algorithm}\footnotesize
\caption{Online Approximate String Matching}
\label{alg:osam}
\begin{algorithmic}[1]
\REQUIRE pattern \textbf{p}, text \textbf{t}, threshold $K$
\ENSURE set of occurrences $\mathcal{S}$
\STATE $l_p = |\mathbf{p}|, i = 0, \mathcal{S} = \emptyset, idx = K, ins = 0, acc = 0, \mathrm{reset}(\mathbf{mem})$ 
\WHILE{true}
\FOR{$l: l_p - K \rightarrow l_p + K$}
\STATE $\mathbf{s} = \mathbf{t}(i;l)$
\STATE $k = \mathrm{lev}(\mathbf{p}, \mathbf{s})$
\IF{\text{R1} is TRUE}
\STATE $idx = k, ins = 1, \mathbf{mem}[idx][:] = [i,l,1]$
\ELSIF{\text{R2} is TRUE or \text{R3} is TRUE}
\STATE $ins = 1, \mathbf{mem}[idx][:] = [i,l,1]$
\ENDIF
\ENDFOR
\IF{$ins == 1$ and $\mathbf{mem}[:][3] == 0$}
\FOR{$j: idx \rightarrow K$}
\IF{$j == idx$ or $\mathbf{mem}[j][3] -acc < \mathbf{mem}[j][2]$}
\STATE $\mathcal{S} = \mathcal{S} \cup \mathbf{t}(\mathbf{mem}[j][1]; \mathbf{mem}[j][2])$
\STATE $acc = \mathbf{mem}[j][2] + acc$
\ENDIF
\ENDFOR
\STATE $idx = K, ins = 0, acc = 0, \mathrm{reset}(\mathbf{mem})$
\ENDIF
\IF{$\mathbf{mem}[idx][3] \neq \mathbf{mem}[idx][2]$}
\FOR{$j: idx \rightarrow K$}
\STATE $\mathbf{mem}[j][3] = \mathbf{mem}[j][3]+1$
\ENDFOR
\ENDIF
\STATE $i=i+1$
\ENDWHILE
\end{algorithmic}
\end{algorithm}

\begin{figure}[!ht]
\centering
\includegraphics[width=0.75\columnwidth, keepaspectratio,trim={0cm 0cm 0cm 0cm},clip]{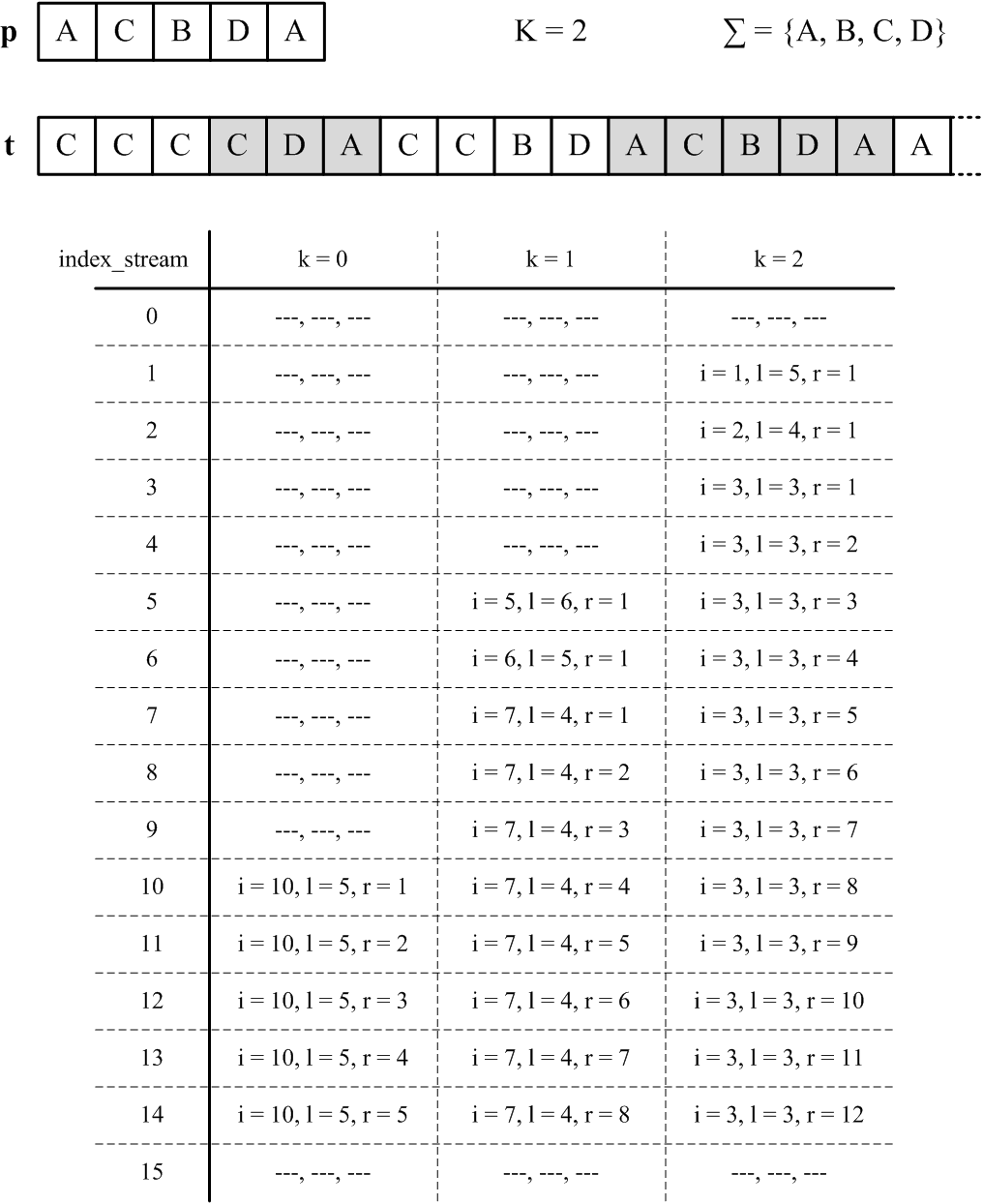}
\caption{Content of \textbf{mem} over time, in the proposed example}
	\label{fig:fig8}
\end{figure}

\paragraph{Example}
The OASM algorithm is executed on the following example, where $\mathbf{p} = \langle ACBDA \rangle$ and $\mathbf{t} = \langle CCCCDACCBDACBDAA \dots \rangle$ are defined over the alphabet $\Sigma = \{A, B, C, D \}$ and $K = 2$. 
Fig. \ref{fig:fig8} illustrates the evolution through time of the variables stored within \textbf{mem}: three occurrences (one per priority) relative to different substrings are overlapping. 
As the counter of the occurrence with priority 0 reaches 5 the validation process is complete. 
The substring $\mathbf{s} = \mathbf{t}[10;5] = \langle ACBDA \rangle$ is the one with highest priority and it represents the occurrence that is validated. 
For the substring $\mathbf{s} = \mathbf{t}[7,4] = \langle CBDA \rangle$, instead, Eq.~\ref{eq:3} has to be evaluated. 
Substituting numerical values in $r(1) - l(0) > l(1)$ gives $8 - 5 > 4$, which is false, so the occurrence with priority 1 is discarded. 
Finally the substring $\mathbf{s} = \mathbf{t}[3,3] = \langle CDA \rangle$ relative to the occurrence with priority 2 has to be added because evaluating $r(2) - l(0) > l(2)$ gives $12 - 5 > 3$, which is true. 
In the last evaluation $l(0)$ is present rather than $l(1)$, since the last validated occurrence is the one with priority 0, whereas the one with priority 1 was discarded. 

\subsection{Complexity Analysis}
The price to pay for filtering out unaccepted occurrences with the proposed validation process is that the validated occurrences are not instantaneously stored in $\mathcal{S}$.\\
To analyze the complexity of the algorithm, we consider the worst case scenario for a generic occurrence with priority $k$. It occurs when (i) overlapping occurrences with all priority values from 0 to $K$ are found, (ii) all the occurrences have maximum length, and (iii) each occurrence with priority $k-1$ begins just before the end of the occurrence with priority $k$.
In this case, the substring corresponding to the occurrence with lowest priority $k_{min}$ cannot be added to $\mathcal{S}$ before $d = l_p (k_{min}+1) + k_{min}(k_{min}-1) / 2$ symbols of $\mathbf{t}$ are processed.\\
The space complexity corresponds to the size of the storage variable \textbf{mem}. According to the discussed worst case scenario, the upperbound for the space required to store \textbf{mem} is $\mathcal{O} \left( l_p(K+1) + \frac{K(K-1)}{2} \left( 2K + 1 \right) \right)$.
\section{Case Study}
\label{sec:casestudy}
microRNAs ({\em miRNAs}) \cite{lee2002microrna,winter2009many} are small non-coding primary transcripted molecules (approx. 22 nucleotides long) which serve as regulators of gene expression and are essential components of normal organism development. 
Found in plants, animals and unicellular eukaryotes, miRNAs control diverse biological functions by promoting degradation or translation inhibition of target messenger RNAs or by carrying out post-transcriptional regulation of gene expression. miRNA genes are dispersed in various genomic locations (intronic, exonic or intergenic regions) and can be transcribed independently or as a part of other host genes.\\
miRNAs are encoded within the genome and are often transcribed by RNA polymerase II or, rarely, by RNA polymerase III as long precursor transcripts (several hundreds or thousands of nucleotides), named primary-microRNAs ({\em pri-miRNAs}) \cite{borchert2006rna,lee2004microrna}. Those pri-miRNAs have the characteristic hairpin (or stem-loop) structure.\\
Within the nucleus, pri-miRNAs are cleaved by the microprocessor complex formed by proteins DROSHA and DGCR8 (also known as PASHA\footnote{PArtner of droSHA}) and the shorter hairpin structure (approx. 72 nucleotides) formed by this cleavage is referred to as precursor-microRNAs ({\em pre-miRNAs}).\\
The pre-miRNAs are transported from the nucleus to the cytoplasm by Exportin-5 together with Ran-Guanine TriPhosphatase \cite{winter2009many}. In the cytoplasm, DICER, a cytoplasmic RNase III type protein, dices the transported pre-miRs near the hairpin loop, yielding a mature miRNA-duplex of approx. 22 nucleotides. The duplex is loaded onto the Argonaute protein ({\em AGO}). One strand of the duplex (the {\em passenger strand}) is discarded, while the other strand (the {\em guide strand} or {\em mature miRNA}) remains in {\em AGO} to form an RNA-Induced Silencing Complex \cite{winter2009many}.\\
The end positions of mature miRNAs are generally assumed as DROSHA cleavage sites. However, pre- and mature-miRNAs are often subject to end modification/processing such as trimming and tailing, which hinders the exact identification of DROSHA cleavage sites \cite{kim2017genome}. Processing of pri-miRNA stem-loops by the DROSHA/DGCR8 complex is the initial step in miRNA maturation and crucial for its function. Nonetheless, the underlying mechanism that determines the DROSHA cleavage site of primary transcripts has remained unclear: while the mechanisms of pre-miRNA recognition and cleavage by DICER are well characterized \cite{park2011dicer}, an understanding of how DROSHA/DGCR8 selectively recognizes and precisely cleaves pri-miRNAs remains unclear \cite{landthaler2004human}.
The problem of identifying cleavage sites for DROSHA relies on the fact that failures at DROSHA processing step might lead to miRNAs down-regulation, a phenomenon observed in cancer patients \cite{thomson2006extensive}.
\subsection{Baseline Algorithms for ASM}
To show the effectiveness of the proposed OASM algorithm, it is compared to two baseline ASM techniques.

The first, referred as {\em Fully Na\"{i}ve}, works in a brute force fashion \cite{hasan2015approximate} according to the following three steps:
\begin{enumerate}
    \item from the input stream, extracts all possible adjacent substrings of any possible length;
    \item evaluates all pairwise Levenshtein distances between substrings and the target sequence;
    \item discards all substrings whose distance with respect to the target sequence is greater than the threshold $K$.
\end{enumerate}
Whilst being the most straightforward method, Fully Na\"{i}ve does not consider multiple shadow hits and all the occurrences found must be stored in order to be filtered in a post-processing stage.  
Further, due to its brute force nature, it is unsuitable for processing large input streams, since the number of possible adjacent substrings of any possible length grows quadratically with the length of the text. Specifically, let $l$ be the length of the text, the number of non-empty adjacent substrings of any possible length is $l(l+1)/2$.
Accordingly, the time and space complexity for evaluating the Levenshtein distance with Wagner-Fischer's algorithm, as described in Sect. \ref{sec:definitions}, is $\mathcal{O}(l \cdot l_p)$ with $l$ and $l_p$ being the input and target string lengths, respectively. 
Thus, the overall complexity in the best case is $\mathcal{O}(l_p)$, whereas in the worst case is $\mathcal{O}(l^2)$. 
Finally, the filtering procedure must scan all possible substrings generated in step 1, and its complexity is again $\mathcal{O}(l^2)$.
By strictly following the three steps above, step 1 might lead to early out-of-memory errors due to the storage of $\mathcal{O}(l^2)$ possible substrings. 
To avoid this, for each extracted substring one can evaluate the distance with respect to the target and discard the substring if the distance is above the user-defined threshold.

The second procedure, referred as {\em Less Na\"{i}ve}, tweaks the Levenshtein matrix $\mathbf{C}$ \cite{navarro2001guided,utan2010fpga}. Again, let $l$ and $l_p$ be the length of the input stream and target string, respectively. 
According to Sect. \ref{sec:definitions}, to keep track of the prefixes one initializes the matrix $\mathbf{C}$ with a dummy row ($0$ to $l$) and a dummy column ($0$ to $l_p$), as shown in Fig. \ref{fig:fig1}. 
Instead, if the dummy row is initialized with all zeros, the Wagner-Fischer's algorithm computes the Levenshtein distance between the target and substrings in the input stream.
Therefore, rather than picking the bottom-right element, one takes all positions in the last row that are below the threshold $K$ as the ending positions of the hits.

\subsection{Qualitative Analysis of the Results}
\begin{figure}[!ht]
\centering
\resizebox{0.8\columnwidth}{!}{%
\includegraphics{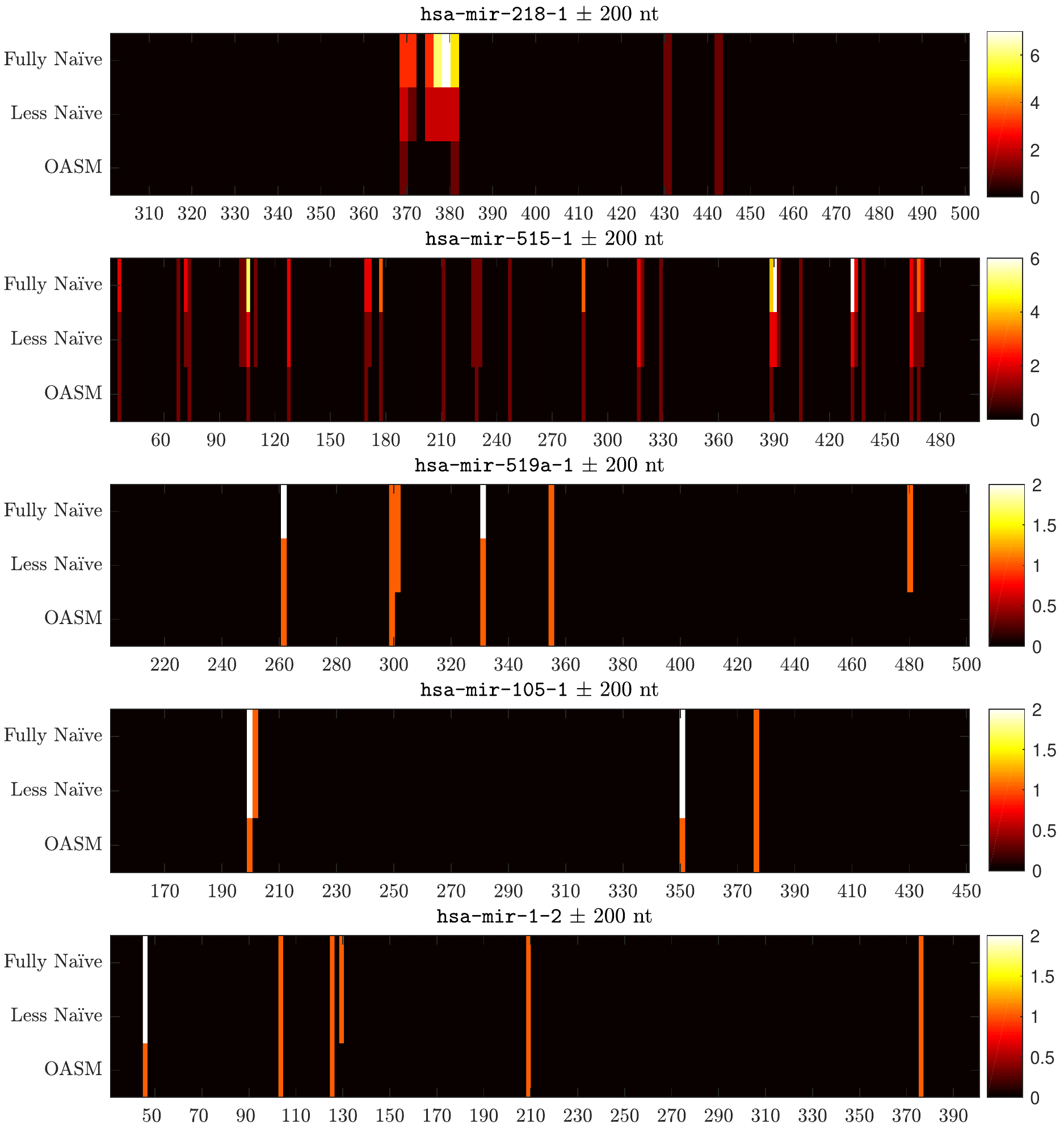}
}
\caption{Hits positions for experiments in Table \ref{tab:tab_effectiveness}. The colour map indicates the number of overlapping occurrences found in a given position.}
\label{fig:figureHits}
\end{figure}

After consultation with field-experts\footnote{Dr. Giulia Piaggio and Dr. Aymone Gurtner. Regina Elena Institute for Cancer Research. Rome, Italy.}, several human pre-miRNAs of interest have been selected for the analysis. 
As discussed above, since the end positions of mature miRNAs cannot be considered as reliable cleavage sites, mining pre-miRNAs neighbourhood regions (both upstream and downstream) can give some further insights to biologists.\\
To evaluate the capability of the proposed algorithm to filter shadow-hits on the fly, five of the suggested pre-miRNAs have been considered, along with the corresponding chromosomes. All data are freely available on ad-hoc biological online databases: the human miRNAs can be found at miRBase \cite{kozomara2013mirbase}, whereas the human chromosomes (assembly GRCh38) can be retrieved from GenomeBrowser \cite{kent2002human}.
The three algorithms, {\em Fully Na\"{i}ve}, {\em Less Na\"{i}ve}, and OASM  have been evaluated on the same data using the same inexactness threshold $K=2$. 
The number of hits and their positions are reported in Table \ref{tab:tab_effectiveness} and  Fig. \ref{fig:figureHits} visually depicts the amount of overlapping occurrences found in each position.
\bgroup
\def\arraystretch{0.95} 
\setlength\tabcolsep{.4em} 
\begin{table}[!ht]
\footnotesize
\centering
\caption{Number of hits across 5 different pre-miRNAs by considering 200 nucleotides neighbourhood (both upstream and downstream). The total number of nucleotides is shown in brackets.}
\label{tab:tab_effectiveness}
\begin{tabular}{ccccc}
\cmidrule[1.5pt]{1-5}
 \textbf{Input Stream} & \textbf{Target} & \textbf{Fully Na\"{i}ve} & \textbf{Less Na\"{i}ve} & \textbf{OASM}\\
\cmidrule[0.5pt]{1-5}
\texttt{hsa-mir-218-1} $\pm$ 200 nt (510 nt)     & \texttt{aaaaaaaa} & 29 & 13 & 4\\
\texttt{hsa-mir-515-1} $\pm$ 200 nt (483 nt)     & \texttt{gcaacc}   & 64 & 39 & 19\\
\texttt{hsa-mir-519a-1} $\pm$ 200 nt (485 nt)    & \texttt{acgttgca} & 8 & 6 & 4\\
\texttt{hsa-mir-105-1} $\pm$ 200 nt (481 nt)     & \texttt{aaccttgg} & 6 & 6 & 3\\
\texttt{hsa-mir-1-2} $\pm$ 200 nt (485 nt)       & \texttt{ctcattca} & 7 & 7 & 5\\
\cmidrule[1.5pt]{1-5}
\end{tabular}
\end{table}
\egroup
From Table \ref{tab:tab_effectiveness} and Fig. \ref{fig:figureHits} clearly emerges that OASM detects each occurrence accurately just one time since, in contrast to the other two algorithms, it is able to discard those detected sequences that are overlapping.
Indeed, neither Fully Na\"{i}ve nor Less Na\"{i}ve can deal with shadow hits, thus requiring a mandatory additional post-processing phase to validate the occurrences. 
Such a post-process results in additional computation and, most importantly, contrarily to the proposed OASM algorithm, the detection cannot be done online.

Tab.~\ref{tab:sw_times} summarizes the time complexity of the ASM algorithms.
Execution times are not reported since the post-processing filtering operation, which is required when using Fully Na\"{i}ve and Less Na\"{i}ve algorithms, is not univocally defined.
In fact, it can be implemented in several ways and executed at different times, in accordance to the requirements of the task at hand. 
A detailed discussion and analysis of the offline filtering operations for the na\"{i}ve algorithms is beyond the scope of this work.

\bgroup
\def\arraystretch{1.5} 
\setlength\tabcolsep{1em} 
\begin{table}[!ht]
\footnotesize
\centering
\caption{Computational complexities of the ASM algorithms. Herein, let $l$ and $l_p$ be the length of the input stream and the length of the pattern to be search within the input stream, respectively.}
\label{tab:sw_times}
\begin{tabular}{lc}
\cmidrule[1.5pt]{1-2}
\textbf{Algorithm} & \textbf{Time Complexity}\\
\cmidrule[0.5pt]{1-2}
Fully Na\"{i}ve & $\mathcal{O}(l^2) \cdot [ \mathcal{O}(l_p)\rightarrow\mathcal{O}(l^2) ]+\mathcal{O}(l^2)$ + offline filtering cost\\
Less Na\"{i}ve & $\mathcal{O}(l \cdot l_p)$ + offline filtering cost\\
OASM & $\mathcal{O}\left( l \cdot \left( l_p (K+1) + \frac{K(K-1)}{2} \right) \right)$\\
\cmidrule[1.5pt]{1-2}
\end{tabular}
\end{table}
\egroup

\section{Proposed Hardware Implementation}
\label{sec:hardware}

\subsection{Top-level instance: \emph{LEV CORE} }

\begin{figure}[!ht]
\centering
\includegraphics[width=0.6\columnwidth, keepaspectratio,trim={0cm 0cm 0cm 0cm},clip]{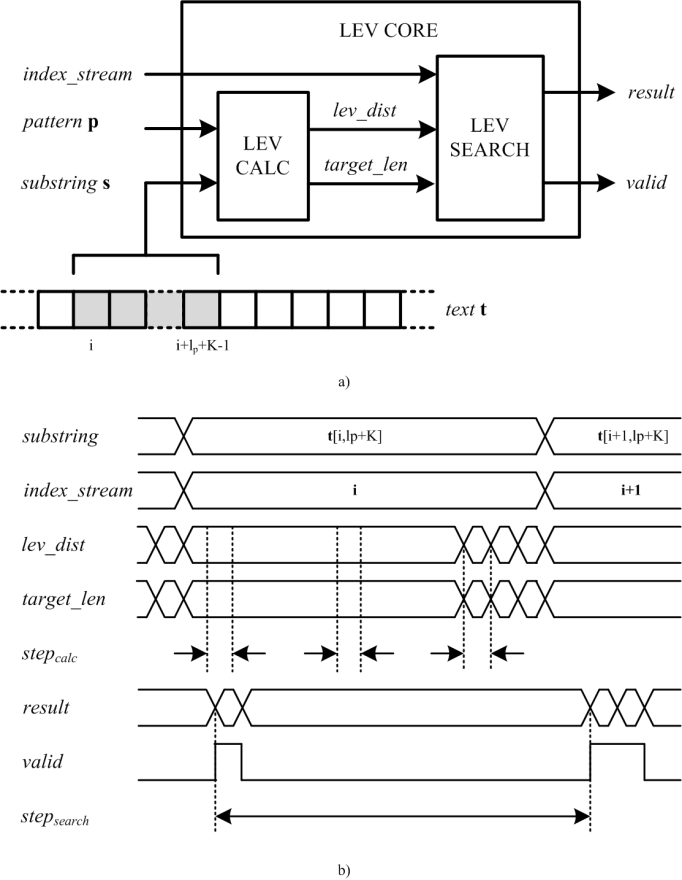}
\caption{Conceptual block scheme for LEV CORE module (a) and relative timing for the main signals (b).}
	\label{fig:fig2}
\end{figure}

The LEV CORE module computes the multiple Levenshtein distances between the \textit{pattern} \textbf{p} and all the \textit{substrings} \textbf{s} extracted from a stream of symbols \textbf{t} and returns only the list of occurrences \textit{result} below a threshold $K$, according to the algorithm described in Sec.~\ref{sec:algo}. 
Fig.~\ref{fig:fig2}(a) depicts the main I/O ports and the two main sub-modules of the LEV CORE:

\begin{itemize}
    \item LEV CALC computes at every new \textit{index\_stream} value the distances between \textbf{p} and the $2K+1$ substrings $\mathbf{s} \in \{ \mathbf{t}[i, l_p-K], \dots, \mathbf{t}[i, l_p+K]$ \}.
    \item LEV SEARCH elaborates the received data to search for all the possible occurrences of \textbf{p} and to validate the found occurrences that satisfy the priority scheme described in Sec.~\ref{sec:algo}, discarding undesired shadow hits.
\end{itemize}

After the pattern \textbf{p} is loaded into the proper registers, the elaboration starts as the continuous stream of data \textbf{t} flows into the module. 
In Fig.~\ref{fig:fig2}(b) a time diagram of the main signals (port-level and internal) is shown.

\subsection{Distance computation: \emph{LEV CALC} }

\begin{figure}[!ht]
\centering
\includegraphics[width=0.8\columnwidth, keepaspectratio,trim={0cm 0cm 0cm 0cm},clip]{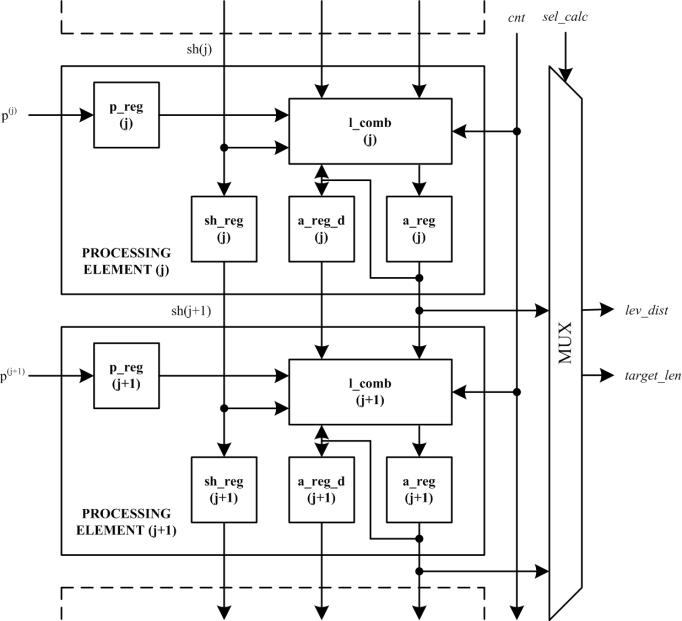}
\caption{Details of the systolic architecture of LEV CALC module.}
	\label{fig:fig3}
\end{figure}

The systolic nature of the DP algorithm used to compute the Levenshtein distance is highly parallelizable and particularly suitable for direct hardware implementation. 
A systolic architecture~\cite{kung1980algorithms} consists of identical processing elements arranged in an array that process data synchronously, executing a short invariant sequence of instructions without an intervening memory to store and exchange results across the pipelined array. 
The advantage of this architecture is to make both the space and time of the calculation of the Levenshtein matrix linear by parallelizing independent processes. 
In the ordinary systolic architectures used in string matching applications~\cite{lipton1985systolic}, two strings are shifted on each other to compare step-by-step each pair of symbols. 
To compute the distance, each symbol needs to be associated to an additional (stored) information, which is the number that identifies the position of the symbol within the stream \textbf{t}. 
The LEV CALC sub-module here introduced implements a counter-based systolic architecture that avoids the storage of \textit{a-priori} known data. 
It computes the Levenshtein distances in a wave-front fashion returning all the elements of the current anti-diagonal of \textbf{C}, filling the matrix in $2 l_p + K - 1$ steps, each one from now on referred as step\textsubscript{calc} (Fig.~\ref{fig:fig2}(b)).
Fig.~\ref{fig:fig3} illustrates the conceptual architecture of two consecutive processing elements of the systolic array, on which the LEV CALC sub-module is based on. 

The generic $j$-th processing element contains both sequential and combinational logic. 
The register $\mathbf{p\_reg}(j)$ is configured with the $j$-th symbol $p^{(j)}$ of the pattern \textbf{p} (if unused, filled with special symbol $\$_1$). 
The register $\mathbf{sh\_reg}(j)$ contains the shifting symbols sh$(j)$ of the substring $\mathbf{s} = \mathbf{t}[i,l_p+K]$ and at each step\textsubscript{calc}, contains a different symbol of \textbf{s}. 
The special symbol $\$_2$ is used to fill the unused \textbf{sh\_reg} positions during the shifting. 
The special symbols $\$_1$ and $\$_2$ belong to the alphabet $\Sigma$ but cannot be used both in \textbf{p} and in \textbf{t}. 
The register $\mathbf{a\_reg}(j)$ stores the result of the $j$-th processing element produced by $\mathbf{l\_comb}(j)$ during the current step\textsubscript{calc}. 
The register $\mathbf{a\_reg\_d}(j)$ represents a delayed version of $\mathbf{a\_reg}(j)$. 
The combinational block $\mathbf{l\_comb}(j)$ elaborates both the information contained in the above mentioned registers and the state of the upward counter \textit{cnt} to yield the $j$-th element of the anti-diagonal of the matrix \textbf{C}.

The counter is used both to allow the shifting process of the symbols sh$(j)$ through the $j$-th processing block, and to provide the suitable $n$ or $m$ values of Eq.~\ref{eq:lev_mat} to $\mathbf{l\_comb}(j)$ block when the corresponding processing element represents an element $c_{n,0}$ or $c_{0,m}$ of \textbf{C} respectively. 
The $j$-th element of the anti-diagonal in the current step\textsubscript{calc} implementing a rearranged, yet equivalent, version of Eq.~\ref{eq:lev_mat} is computed as
\begin{equation}
\label{eq:lcomb}
    \mathbf{l\_comb}(j) = \mathrm{min}
    \begin{bmatrix}
    c_\mathrm{left}(j), \\
    c_\mathrm{upper}(j), \\
    c_\mathrm{upper-left}(j) \\
    \end{bmatrix}
    + \phi (j),
\end{equation}
where
\[
c_\mathrm{left}(j) = 
\begin{cases}
\mathbf{a\_reg\_d}(j), & \text{if}\; j < cnt \\
cnt, & \text{otherwise}
\end{cases}
\]
\[
c_\mathrm{upper}(j) = 
\begin{cases}
cnt, & \text{if}\; j=0 \\
\mathbf{a\_reg}(j-1), & \text{if}\; j > 0 \\
\end{cases}
\]
\[
c_\mathrm{upper-left}(j) = 
\begin{cases}
\mathbf{a\_reg\_d}(j-1), & \text{if}\; j < cnt \\
cnt-1, & \text{otherwise}
\end{cases}
\]
\[
\phi(j) = 
\begin{cases}
1, \;\text{if}\; c_\mathrm{upper-left}(j) > \mathrm{min}
    \begin{bmatrix}
    c_\mathrm{left}(j), \\
    c_\mathrm{upper}(j), \\
    c_\mathrm{upper-left}(j) \\
    \end{bmatrix} \\
\mathbf{p\_reg}(j) \neq \mathbf{sh\_reg}(j), \;\text{otherwise}.    
\end{cases}
\]

\begin{figure}[!ht]
\centering
\includegraphics[width=0.6\columnwidth, keepaspectratio,trim={0cm 0cm 0cm 0cm},clip]{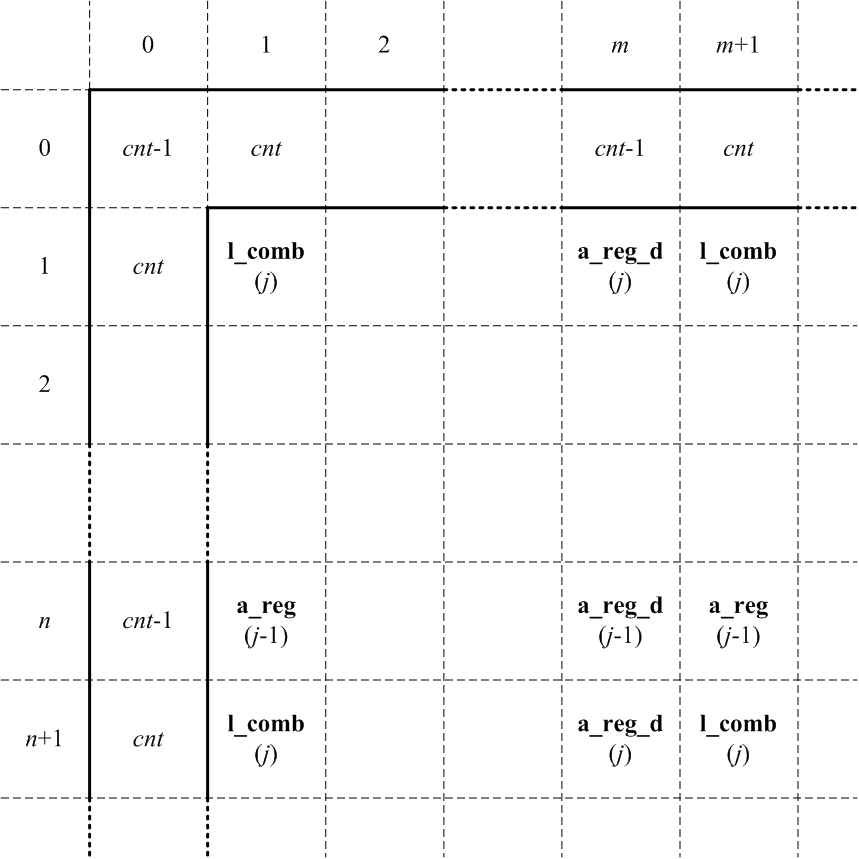}
\caption{Representation of the neighbors in the C matrix to compute $\mathbf{l\_comb}(j)$.}
	\label{fig:fig4}
\end{figure}

In Fig.~\ref{fig:fig4} the four possible combinations of neighbors for the computation of $\mathbf{l\_comb}(j)$ are located in the \textbf{C} matrix frame to show how Eq.~\ref{eq:lev_mat} turns into Eq.~\ref{eq:lcomb}.

\begin{figure}[!ht]
\centering
\includegraphics[width=0.7\columnwidth, keepaspectratio,trim={0cm 0cm 0cm 0cm},clip]{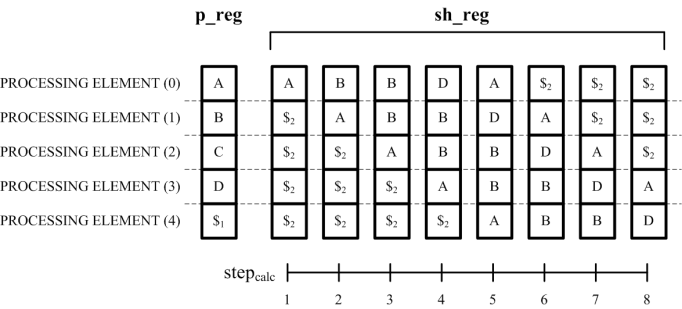}
\caption{Evolution of the content of \textbf{sh\_reg} step by step to implement the string shifting.}
	\label{fig:fig5}
\end{figure}

Finally, the MUX selector \textit{sel\_calc} is constant and uniquely determined with the length of the pattern \textbf{p} and the value of the threshold $K$. 
Fig.~\ref{fig:fig5} shows the content of \textbf{p\_reg} and \textbf{sh\_reg} in the inherent shifting mechanism for the example discussed in Sec.~\ref{sec:definitions}.

\subsection{Search and validation: \emph{LEV SEARCH} }

\begin{figure}[!ht]
\centering
\includegraphics[width=0.75\columnwidth, keepaspectratio,trim={0cm 0cm 0cm 0cm},clip]{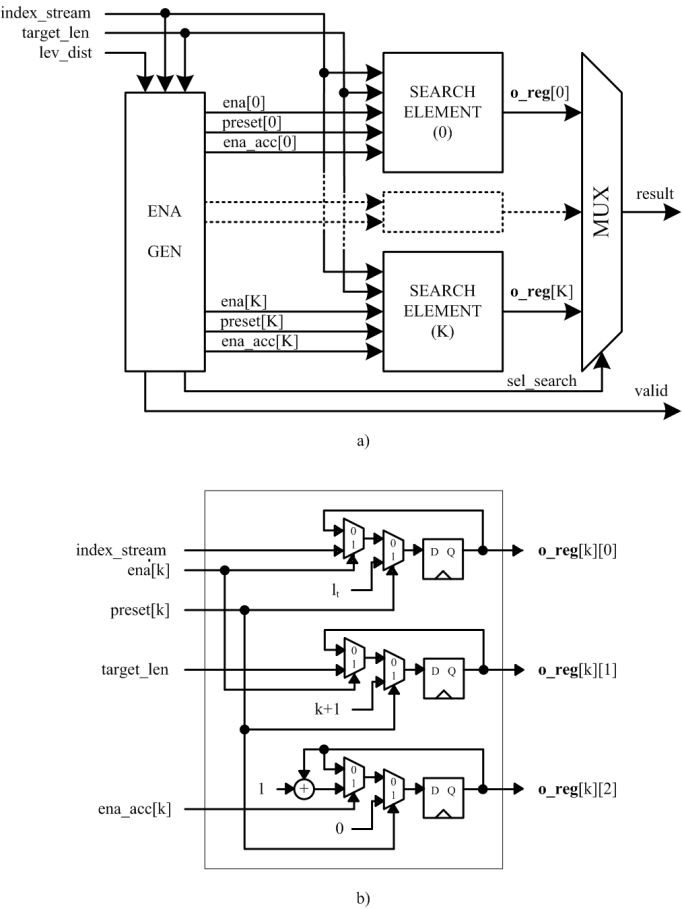}
\caption{Conceptual scheme of the LEV SEARCH architecture (a) and detail of the $k$-th SEARCH ELEMENT sub-block (b)}
	\label{fig:fig6}
\end{figure}

The outputs generated by LEV CALC, together with \textit{index\_stream}, represent the inputs of the LEV SEARCH sub-module whose architecture is schematized in Fig.~\ref{fig:fig6}(a). 
This module implements the online search algorithm described in Sec.~\ref{sec:algo} and formalised in Alg.~\ref{alg:osam}.

In an infinite loop cycling on $i$ (corresponding to a new \textit{index\_stream}) the Levenshtein distances between \textbf{p} and all the possible $2K+1$ substrings $\mathbf{s} \in \{ \mathbf{t}[i,l_p-K], \dots, t[i,l_p+K] \}$ returned by LEV CALC, which satisfy both the described priority rules (R1, R2, and R3) and validation process, are stored in $\mathcal{S}$. 
The logic relative to the various checks performed by the algorithm is located into the ENA GEN sub-block.
There, $[i,l,k]$ = [\textit{index\_stream}, \textit{target\_len}, \textit{lev\_dist}] are processed by the priority rules (see Sec.~\ref{sec:algo}), and Eq.~\ref{eq:3} is evaluated on the values of the counters $r(k)$ during the validation process. 
Within the same block, the variables \textit{idx}, \textit{ins}, and \textit{acc} defined in Alg.~\ref{alg:osam} are implemented as a status register, a flag and an accumulation register, respectively, and are used to generate the signals ena, preset, ena\_acc, and valid.
Fig.~\ref{fig:fig6}(b) depicts a detail of the sequential section which represents the state of the algorithm (SEARCH ELEMENTs). 
The two-dimensional array \textbf{o\_reg} of size $(K+1) \times 3$ implements the storage variable \textbf{mem} in Alg.~\ref{alg:osam} and it allows to retrieve information relative to the current eligible occurrences.
The procedure executed in LEV CALC consists in a constant number of step\textsubscript{calc}, whereas the algorithm implemented in LEV SEARCH may have a different duration depending on the validation process that is executed only if eligible occurrences are found. 
A LEV SEARCH cycle is called here step\textsubscript{search} and its duration coincides with number of step\textsubscript{calc} necessary to complete on LEV CALC cycle multiplied by its duration
\begin{equation}
    \label{eq:6}
    T_{\text{step}_\text{search}} = (2 \cdot l_p+K-1)  T_{\text{step}_\text{calc}}.
\end{equation}
The proposed hardware architecture executes one step\textsubscript{calc} in one clock cycle ($T_\text{clk}$), so the LEV CALC block generates the first \textit{lev\_dist} sample after $2 l_p-K-1$  clock cycles and in the same cycle the LEV SEARCH block starts processing it:
\[
\text{lat}_\text{calc} = 2 l_p+K-1
\]
\[
\text{lat}_\text{search} = 
\begin{cases}
K + 1 & \text{without validated occurrences} \\
2K + 1 & \text{with validated occurrences} \\
\end{cases}
\]

\begin{figure}[!ht]
\centering
\includegraphics[width=0.7\columnwidth, keepaspectratio,trim={0cm 0cm 0cm 0cm},clip]{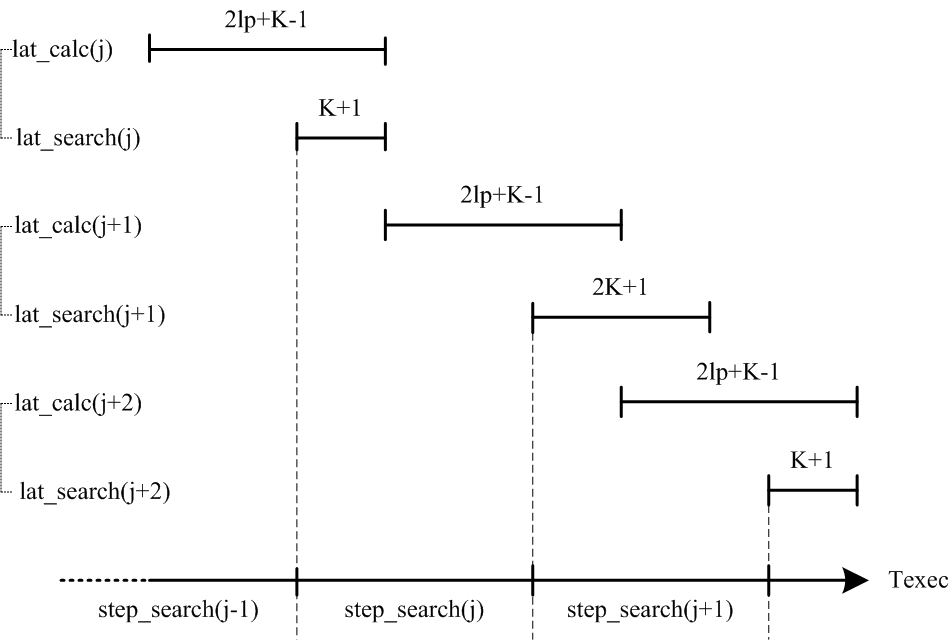}
\caption{Time execution diagram in the pipelined structure}
	\label{fig:fig7}
\end{figure}

Although the latency of the LEV SEARCH is data dependent, the operations of the two modules can be pipelined.
The total execution time depends on the latency of the two blocks:
\begin{equation}
    \label{eq:7}
    T_\mathrm{exec} \approx T_{\text{step}_\text{search}} l_t = (2l_p + K-1) \cdot T_\text{clk}l_t
\end{equation}
where $T_\text{clk}$ is the clock period. 
It should be noticed that Eq.~\ref{eq:7} is valid when $K < 2 (l_p -1)$, which is always verified since the threshold $K$ is always lower than $l_p$. 
Fig.~\ref{fig:fig7} depicts the pipelining process.

\subsection{Resource usage and parallel search}
\begin{figure}[!ht]
\centering
\includegraphics[width=0.45\columnwidth, keepaspectratio,trim={0cm 0cm 0cm 0cm},clip]{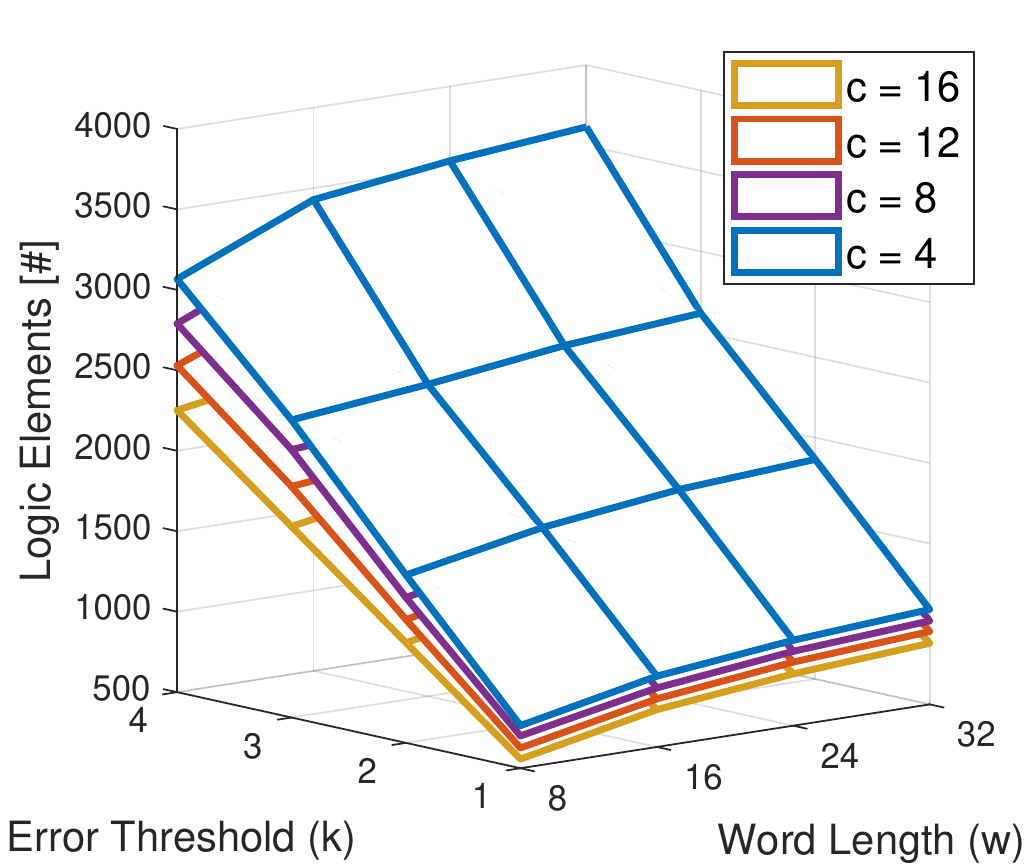}
\caption{Linear relationship between design parameters and number of LEs}
	\label{fig:fig9}
\end{figure}

The LEV CORE has been completely described in VHDL and parameterized in terms of
\begin{itemize}
    \item $l_\text{p\_max}$: maximum number of symbols per pattern;
    \item $l_\text{symb}$: number of bits per symbol;
    \item $K$: maximum threshold.
\end{itemize}
Tab.~\ref{tab:tab1} reports synthesis results in terms of Logic Elements (LE{[}\#{]}) used, when varying the design parameters.
\begin{table}[!ht]
\centering
\footnotesize
\caption{LE utilization relationship with design parameters}
\label{tab:tab1}
\begin{tabular}{llll|llll}
\cmidrule[1.5pt]{1-8}
$l_\text{symb}$     & $l_\text{p\_max}$   & $K$ & LE{[}\#{]} & $l_\text{symb}$     & $l_\text{p\_max}$   & $K$ & LE{[}\#{]} \\
\cmidrule[1.5pt]{1-8}
\multirow{16}{*}{4} & \multirow{4}{*}{8}  & 1   & 557        & \multirow{16}{*}{8} & \multirow{4}{*}{8}  & 1   & 626        \\
                    &                     & 2   & 732        &                     &                     & 2   & 799        \\
                    &                     & 3   & 822        &                     &                     & 3   & 895        \\
                    &                     & 4   & 880        &                     &                     & 4   & 953        \\
                    & \multirow{4}{*}{16} & 1   & 1122       &                     & \multirow{4}{*}{16} & 1   & 1266       \\
                    &                     & 2   & 1286       &                     &                     & 2   & 1443       \\
                    &                     & 3   & 1391       &                     &                     & 3   & 1536       \\
                    &                     & 4   & 1472       &                     &                     & 4   & 1610       \\
                    & \multirow{4}{*}{24} & 1   & 1687       &                     & \multirow{4}{*}{24} & 1   & 1936       \\
                    &                     & 2   & 1840       &                     &                     & 2   & 2025       \\
                    &                     & 3   & 1960       &                     &                     & 3   & 2151       \\
                    &                     & 4   & 2064       &                     &                     & 4   & 2217       \\
                    & \multirow{4}{*}{32} & 1   & 2252       &                     & \multirow{4}{*}{32} & 1   & 2530       \\
                    &                     & 2   & 2394       &                     &                     & 2   & 2869       \\
                    &                     & 3   & 2529       &                     &                     & 3   & 3002       \\
                    &                     & 4   & 2656       &                     &                     & 4   & 3068      \\
\cmidrule[1.5pt]{1-8}
$l_\text{symb}$      & $l_\text{p\_max}$   & $K$ & LE{[}\#{]} & $l_\text{symb}$      & $l_\text{p\_max}$   & $K$ & LE{[}\#{]} \\
\cmidrule[1.5pt]{1-8}
\multirow{16}{*}{12} & \multirow{4}{*}{8}  & 1   & 701        & \multirow{16}{*}{16} & \multirow{4}{*}{8}  & 1   & 765        \\
                     &                     & 2   & 867        &                      &                     & 2   & 940        \\
                     &                     & 3   & 962        &                      &                     & 3   & 1031       \\
                     &                     & 4   & 1019       &                      &                     & 4   & 1090       \\
                     & \multirow{4}{*}{16} & 1   & 1403       &                      & \multirow{4}{*}{16} & 1   & 1543       \\
                     &                     & 2   & 1579       &                      &                     & 2   & 1706       \\
                     &                     & 3   & 1670       &                      &                     & 3   & 1810       \\
                     &                     & 4   & 1738       &                      &                     & 4   & 1865       \\
                     & \multirow{4}{*}{24} & 1   & 2165       &                      & \multirow{4}{*}{24} & 1   & 2349       \\
                     &                     & 2   & 2233       &                      &                     & 2   & 2438       \\
                     &                     & 3   & 2352       &                      &                     & 3   & 2547       \\
                     &                     & 4   & 2402       &                      &                     & 4   & 2616       \\
                     & \multirow{4}{*}{32} & 1   & 2790       &                      & \multirow{4}{*}{32} & 1   & 3067       \\
                     &                     & 2   & 3163       &                      &                     & 2   & 3428       \\
                     &                     & 3   & 3269       &                      &                     & 3   & 3537       \\
                     &                     & 4   & 3359       &                      &                     & 4   & 3616      \\
\cmidrule[1.5pt]{1-8}
\end{tabular}
\end{table}
%
The largest synthesized LEV CORE modules ($l_\text{symb} = 16$, $l_\text{p\_max} = 32$, $K = 4$) uses only 3\% of the total available resources on Altera Cyclone IV E FPGA (114800 LE). 
Fig.~\ref{fig:fig9} depicts the results of Tab.~\ref{tab:tab1}, providing a clearer view of the resource usage as the design parameters vary. 
The planar shape of the surfaces and the lack of intersections shows the linear relation between resource usage and design parameters.

\begin{figure}[!ht]
\centering
\includegraphics[width=0.6\columnwidth, keepaspectratio,trim={0cm 0cm 0cm 0cm},clip]{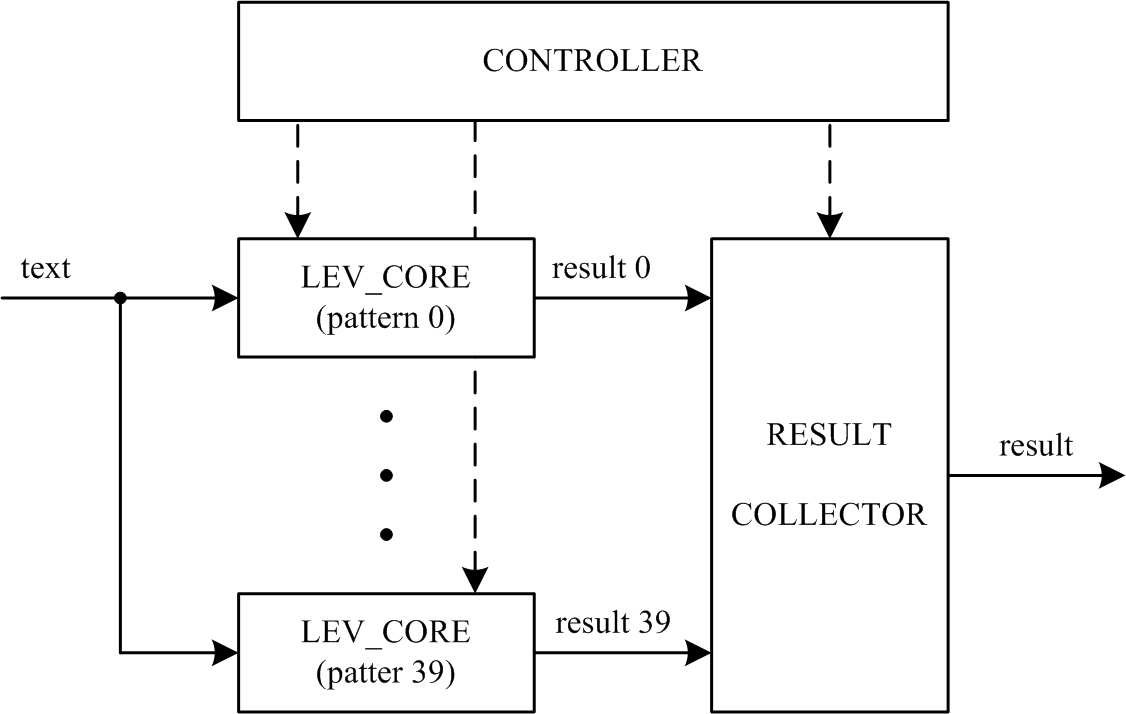}
\caption{Conceptual scheme of a MOASM system}
	\label{fig:MOASM}
\end{figure}

A parallel search of different patterns over the same text \textbf{t} can be done efficiently, by deploying multiple instances of LEV CORE.
This is implemented by the Multiple OASM (MOASM) system, whose architecture is schematized in Fig.~\ref{fig:MOASM}.
The synthesis of a MOASM system with 40 LEV CORE modules ($l_\text{symb} = 8$, $l_\text{p\_max} = 16$, $K = 4$) and additional control logic resulted in 62936 LE, which corresponds to the 55\% of the resource usage on the same target FPGA.

\section{Experiments}\label{sec:experiments}
In this section are reported experiments on synthetic data in order to compare the speed performances of the software (SW-OASM) and hardware (HW-OASM) implementation of the proposed OASM algorithm. 
SW-OASM has been implemented in \texttt{C++} and executed on a CPU Intel Core i7-4700MQ @ 2.40 GHz. 
HW-OASM has been described in VHDL-1993 and implemented on the FPGA Altera Cyclone IV E mounted on the board Terasic DE2-115. 
Two tests are performed using the same setup with randomly generated sequences \textbf{t} of $l_t = 3104$ symbols defined over the alphabet $\Sigma = \{A,B,C,D\}$, and randomly generated patterns \textbf{p} that are processed by both the implementations, for different values of patterns length $l_p$ and threshold $K$.

\begin{figure}[!ht]
\centering
\includegraphics[width=0.6\columnwidth, keepaspectratio,trim={0cm 0cm 0cm 0cm},clip]{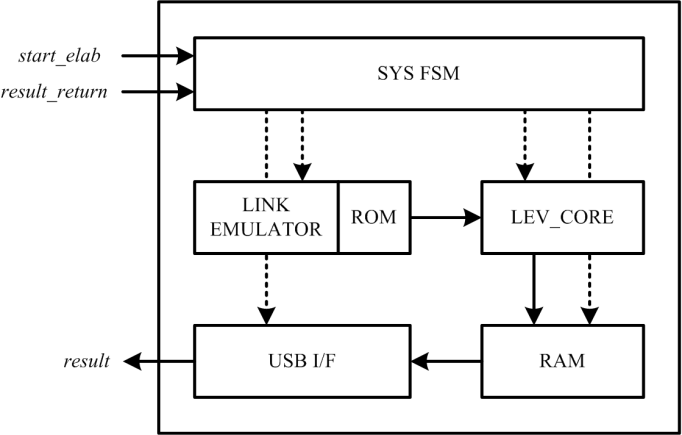}
\caption{System implemented for executing the experiments}
	\label{fig:fig10}
\end{figure}

Recent FPGAs commonly host a high speed link (e.g. PCIe 2.0 with 5 GT/s and the per-lane throughput 500 MB/s) so, to compensate for the lack of a high speed link on the available FPGA, a simple link emulator for the online input data-transfer has been designed. 
This solution can be considered acceptable since the purpose of the tests is measuring the processing time of the LEV CORE block. 
Since, two special symbols $\$_1$ and $\$_2$ are added to the alphabet $\Sigma$, each symbol in \textbf{p} and \textbf{t} is coded with $l_\text{symb} = 3$ bits.
The text \textbf{t} is stored into the FPGA in 1 Mbit embedded memory and is sent, symbol by symbol, to the processing logic that operates at 100 MHz (300 Mbit/s). 
The resulting data flow is stored into another 1 Mbit embedded memory element and is sent to a PC by a low-speed USB link, to be analyzed once the test is completed. 
Fig.~\ref{fig:fig10} schematises the HW-OASM architecture implemented for the tests, whose details are the following.
\begin{itemize}
    \item LEV CORE: processing block with hardwired pattern \textbf{p}, maximum pattern length $l_p = 15$, alphabet size = 3 bit (extended due to special symbols), maximum threshold $K = 5$.
    \item LINK EMULATOR: 1 Mbit ROM, containing the text \textbf{t} (5 symbols per memory location) plus control logic. It sends a new symbol to the LEV CORE block at every new incoming request pulse.
    \item RAM: one 1 Mbit RAM to dump the results (1 Mbit), 16 bit index (sufficient to recognise an occurrence), 3 bit threshold, 5 bit length (max. detected length is $l_p + K = 20$).
    \item USB I/F: USB device interface to transfer the dumped results to an external PC.
    \item SYS FSM: finite state machine to coordinate the I/O operations. Main control signals are \textit{start\_elab} (start of elaboration) and \textit{result\_return} (command to start returning the data stored in RAM via USB I/F).
\end{itemize}

\begin{figure}[!ht]
\centering
\includegraphics[width=0.4\columnwidth, keepaspectratio,trim={0cm 0cm 0cm 0cm},clip]{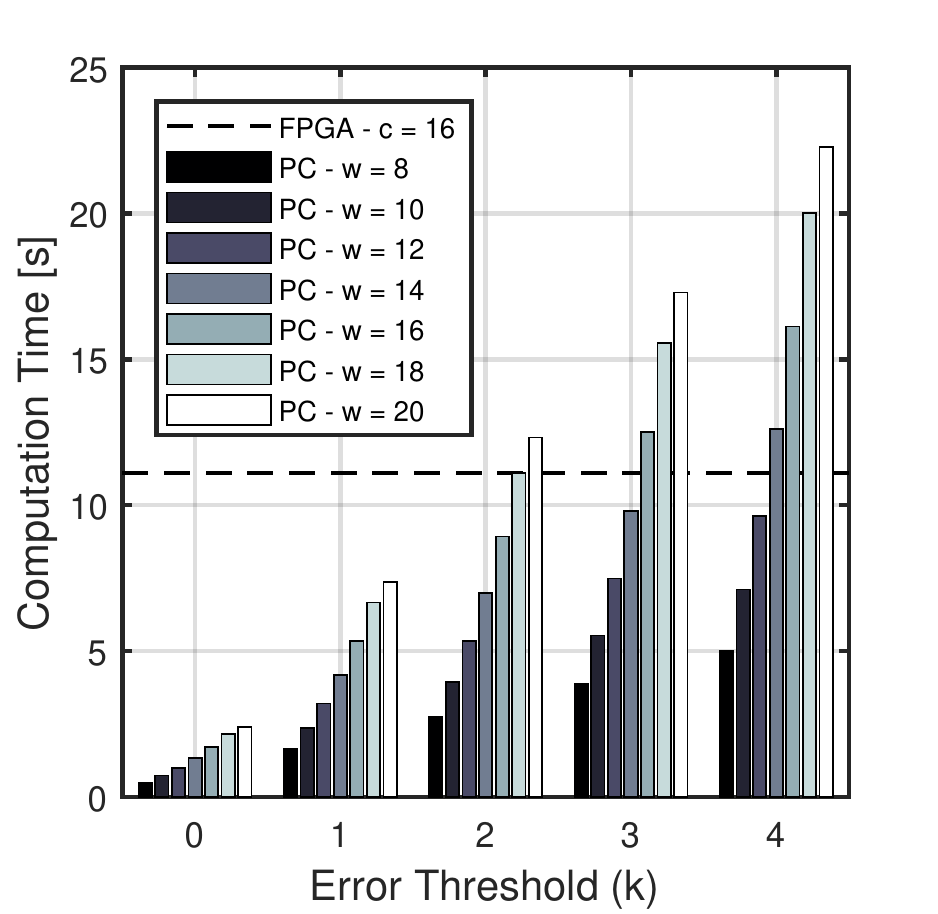}
\caption{SW-OASM computation time for different word length (w) and error threshold (k) pairs. HW-OASM computation time (dashed) is constant on the same set of parameters used for SW-OASM}
	\label{fig:fig11}
\end{figure}

The wall-clock time of the SW-OASM execution is not deterministic, since there are usually other processes running in the background that consume computational resources.
Therefore, each test run with SW-OASM for a specific configuration has been repeated 100 times with independently randomly generated datasets and the results are reported as mean and standard deviation. 
Regarding the tests run with HW-OASM, the results prove the effectiveness of the linear deterministic model for the execution time computation described in Eq.~\ref{eq:7} with $T_\text{clk} = 10 ns$. 
The experimental results slightly depart from the theoretical ones because of implementation details that prevent $T_{\text{step}_\text{search}}$ in Eq.~\ref{eq:6} to be just dependent on $l_p$ and $K$. 
In Fig.~\ref{fig:fig11} the speed performance are reported, when different thresholds $K$ are used.

Two different tests are performed. 
In the first, whose results are in Tab.~\ref{tab:tab2}, random patterns with lengths $l_p = \{5, 7, 10, 15 \}$ are generated and a threshold $K=3$ is used.
\bgroup
\def\arraystretch{0.95} 
\setlength\tabcolsep{.4em} 
\begin{table}[!ht]
\centering
\footnotesize
\caption{Speed performance comparison between SW-OASM and HW-OASM implementation for the first test}
\label{tab:tab2}
\begin{tabular}{ccc}
\cmidrule[1.5pt]{1-3}
 $\mathbf{l_p}$ & \textbf{SW-OASM [s]} & \textbf{HW-OASM [s]} \\
\cmidrule[0.5pt]{1-3}
5 &	    46.51$\pm$0.74	& 0.0039 \\
7 &	    59.991$\pm$1.42	& 0.0051 \\
10 &	78.973$\pm$3.54	& 0.0069 \\
15 &	97.004$\pm$1.76	& 0.0099 \\
\cmidrule[1.5pt]{1-3}
\end{tabular}
\end{table}
\egroup

In the second test, with results reported in Tab.~\ref{tab:tab3}, random patterns with fixed length $l_p = 5$ are generated and different thresholds $K=\{2, 3, 4, 5\}$ are used.

\bgroup
\def\arraystretch{0.95} 
\setlength\tabcolsep{.4em} 
\begin{table}[!ht]
\footnotesize
\centering
\caption{Speed performance comparison between SW-OASM and HW-OASM implementation for the second test}
\label{tab:tab3}
\begin{tabular}{ccc}
\cmidrule[1.5pt]{1-3}
 $\mathbf{K}$ & \textbf{SW-OASM [s]} & \textbf{HW-OASM [s]} \\
\cmidrule[0.5pt]{1-3}
2 & 9.731$\pm$0.50	& 0.0096 \\
3 & 15.179$\pm$0.45	& 0.0099 \\
4 & 30.858$\pm$2.90	& 0.0102 \\
5 & 67.953$\pm$3.14 & 0.0105 \\
\cmidrule[1.5pt]{1-3}
\end{tabular}
\end{table}
\egroup
\section{Conclusions}
\label{sec:concl}
Approximate String Matching (ASM) is a particular case of subgraph matching and is a fundamental tool in many practical applications, such as bioinformatics, cybersecurity, diagnostic systems and financial trading. 
However, when using a tolerance threshold, the ASM can produce shadow hits, i.e. it can identify close multiple hits that, in turn, introduce unwanted uncertainty in the overall template matching procedure and thus in the knowledge discovery task. 
Of course, the shadow hits could be filtered out offline afterwards in a post-processing step. 
However, when dealing with online applications an alternative approach must be followed. 

This paper introduces the online ASM (OASM) algorithm to filter out shadow hits online, which is based on a validation procedure that takes advantage of side information provided during Levenshtein distance computations. 
OASM is designed in plain dynamic programming and is characterized by a high degree of parallelism, which can be leveraged to design an efficient FPGA implementation. 
The proposed architecture allows to perform such a complex retrieval procedure relying on inexpensive hardware, such as the adopted entry level Altera Cyclone IV E FPGA, where only 3\% of available logic elements are used. 
This envisage further opportunities to increase the parallelism of the whole procedure, by allocating on the FPGA more LEV CORE units to concurrently search for instances of multiple motifs on the same sequence, or even on different sequences. 
The execution time of the hardware implementation of such an enhanced implementation (MOASM system) is independent of the number of the instantiated LEV CORE modules, whereas the computational time in the software counterpart scales linearly with the number of pattern searched. 
Especially when the MOASM system is implemented on high-end FPGA chips that allow the instantiation of many LEV CORE modules, the I/O interface (usually a PCI-Express bus) may easily become congested, due to the possible huge number of results convoyed on a single communication bus. 
This underlines the importance of the proposed approach to filter out unwanted shadow hits directly on the FPGA, which can greatly reduce the overall I/O throughput. 
An important future work regards the implementation of a complete MOASM system 
and the transfer of the Intellectual Property on a hi-performance FPGA, such as the Stratix V GX.

\bibliographystyle{unsrtnat}  
\bibliography{Biblio}


\end{document}